# Consensus in Software Engineering: A Cognitive Mapping Study

Pontus Johnson
KTH
Stockholm, Sweden
+47 87906825
pontusj@kth.se

Paul Ralph
University of Auckland
Auckland, New Zealand
+64 9 373 7599
paul@paulralph.name

Mathias Ekstedt
KTH
Stockholm, Sweden
+47 87906867
mathias.ekstedt@ics.kth.se

Iaakov Exman
The Jerusalem College of Engineering
Jerusalem, Israel
+972 2 6588000
iaakov@jce.ac.il

Michael Goedicke
paluno, University of Duisburg-Essen
Essen, Germany
+49 201183 3481
michael.goedicke@paluno.uni-due.de

## ABSTRACT
*Background:* Philosophers of science including Collins, Feyerabend, Kuhn and Latour have all emphasized the importance of consensus within scientific communities of practice. Consensus is important for maintaining legitimacy with outsiders, orchestrating future research, developing educational curricula and agreeing industry standards. Low consensus contrastingly undermines a field's reputation and hinders peer review. *Aim:* This paper aims to investigate the degree of consensus within the software engineering academic community concerning members' implicit theories of software engineering. *Method:* A convenience sample of 60 software engineering researchers produced diagrams describing their personal understanding of causal relationships between core software engineering constructs. The diagrams were then analyzed for patterns and clusters. *Results:* At least three schools of thought may be forming; however, their interpretation is unclear since they do not correspond to known divisions within the community (e.g. Agile vs. Plan-Driven methods). Furthermore, over one third of participants do not belong to any cluster. *Conclusion:* Although low consensus is common in social sciences, the rapid pace of innovation observed in software engineering suggests that high consensus is achievable given renewed commitment to empiricism and evidence-based practice.

## Categories and Subject Descriptors
D.2.0 [**Software Engineering**]: General

## 1. INTRODUCTION
Scientific consensus simply refers to the level of general and widespread (but not perfect or unanimous) agreement within a particular scientific community on key topics. Scientific progress is exceptionally difficult in fields that lack consensus. Without consensus, all facts seem equally relevant and "different men confronting the same range of phenomena … describe and interpret them in different ways" [21]. Rather than incrementally improving good theories and rejecting bad theories, academics in low consensus fields talk past each other with successive, disconnected results. Lack of consensus therefore inhibits the development of a cumulative body of knowledge, undermines the reliability of peer review, and delegitimizes the field to outsiders. Educational curricula and industry standards cannot reflect the non-existent consensus.

When consensus is high, however, the scope of the field narrows and fewer types of investigations are perceived as meaningful. The ensuing development of specific norms, specialized equipment and advanced skills highlights previously obscured anomalies, which trigger the next paradigm shift [21]. The pace of innovation increases and more consensus develops as the advancing discovery frontier eclipses bickering over yesterday's disagreements [10]. Scientists focus on extending one another's work rather than arguing.

Meanwhile, much software engineering (SE) research aims to develop technologies and practices that will help developers to succeed in their SE initiatives. However, due to the complexity of software development in practice, demonstrating a direct causal link between a specific artifact and overall performance is extraordinarily difficult. Empirical researchers therefore focus on intermediate variables including *quality of specifications, financial risk* and *developer motivation*. How these intermediate variables interact, however, remains unclear.

Here, consensus is important regardless of its correctness. Problems with clear, widely-accepted theories may be rapidly revealed by empirical research. Clear, consensus theory may therefore be quickly improved or rejected. Unclear theory or general dissensus, contrastingly, cannot be easily tested and rapidly improved. As

Francis Bacon famously said, "Truth emerges more readily from error than from confusion."

Moreover, Kuhn [21] explains that scientific communities mature in three stages – 1) in the "no science at all" stage no common ground exists between the scholars in a discipline; 2) in the "pre-science" stage, a limited number of competing schools hold contradicting theories; 3) in the "normal science" stage, the vast majority adhere to a single theoretical base. This motivates the following research question.

> **Research Question:** *What (Kuhnian) stage of consensus best describes the software engineering academic community, specifically concerning causal relationships between core constructs?*

Here, a *core construct* is a variable that is often hypothesized to affect software engineering outcomes, or is commonly associated with SE performance. Core constructs may be positively or negatively related to outcomes or each other. Moreover, a core construct is something popularly *believed* to influence SE outcomes – whether or not this belief is not correct. More generally, this paper does not address correctness or accuracy in any sense; rather, we focus on the similarities and differences of SE academics' belief structures.

We not turn to the study's empirical methodology (§2), findings (§3) and their interpretation (§4). Next, we discuss related research on scientific consensus (our motivation), causal graphs (our data), previous attempts to generate success theories and how our results relate to general theory in software engineering (§5). Section 6 concludes the paper by summarizing its contributions, limitations and implications for future research.

## 2. METHODOLOGY

The general idea of the study is to have SE academics diagram their mental models of relationships among core SE constructs and then compare the diagrams to establish the extent and nature of consensus. This section describes the pre-study used to generate the list of constructs, population and sampling, the diagramming task, hypotheses, and the analytical approach.

### 2.1. Pre-Study

Practically speaking, it seems unlikely that even SE experts would be capable of generating a comprehensive list of variables quickly and without forethought. We therefore decided to provide a list of variables from which to choose. We developed this list using the following procedure.

1. Search Google Scholar for articles (published in sources including the term *software engineering*) which contain any of the following terms: improve, effect, impact, benefits, increase, effects, achieve, affect, factors, increased, causes, benefit, impacts, influence, affects, cause, caused, influences, lead to, contribute, improves, depends on, increases, determines, contributes, determine, result in, function of, leads to, gain, decrease, gains, accomplish, variables.

2. For each query, examine the first 20 results.

3. Record each causal proposition (e.g. 'developer motivation contributes to project performance').

4. Extract the independent variable (e.g. 'developer motivation').

5. Delete any application-area delimitation (e.g. 'Performance Problems in SMP Application' becomes 'Performance Problems').

6. Rephrase any non-variable concepts as variables (e.g. 'aspect-oriented programming' becomes "use of aspect-oriented programming').

This process produced 180 snippets containing 330 variables. Some variables (e.g. *integration of information retrieval, execution and link analysis algorithms*) appeared irrelevant while others (e.g. *software quality*) appeared to be dimensions, rather than antecedents, of success. Each of the five authors therefore independently organized the list of variables according to the following procedure:

1. If no experts would likely consider a variable a reasonable cause of success, remove it

2. If most experts would likely consider a variable a dimension, rather than cause, of success, remove it

3. Merge closely related variables (e.g. *use of agile methodology* and *use of agile methods*).

Following independent clustering, the authors met and resolved cluster differences by agreement. This produced 28 core constructs (Table 1). As an additional check, we informally asked 24 attendees of the 35th International Conference on Software Engineering to name three or more constructs that are important for software engineering success. All 103 answers corresponded to constructs in the initial list.

### 2.2. Population and Sampling

As the purpose of the paper is to make inferences about consensus within the SE academic community, the population of interest may be stated as active SE academics. Consequently, we interviewed a convenience sample of 60 attendees of the 35th International Conference on Software Engineering (ICSE 2013). We attracted participation by laying out the colorful task materials (below) in visible areas and inviting passing attendees to participate. In practice, many of the participants self-selected by approaching us to ask what we were doing. All five authors participated in the data collection at different locations around the conference areas.

### 2.3. Task Structure and Data Collection

The purpose of the task was to have participants construct a diagram (Figure 1) representing their mental model of how the constructs relate. To help structure this process, we provided a top-level dependent variable—*Software Engineering Success* (SES) [32]—to which the constructs should directly or indirectly relate. To this end, the task proceeded in six steps:

1. The participant picked approximately 10 construct cards, which have the greatest influence on SES.

2. The interviewer spread out the chosen constructs on a table, with the SES card in the middle.

3. The participant used the arrow cards to indicate causal relationships.

4. For each construct card with an arrow pointing in, the interviewer added an *other* card (explained below).

5. The participant indicated the direction and relative strength of each relationship using one number card per relationship. Here, -3 indicates a strong inverse relationship, and +3 indicates a strong direct relationship.

6. The interviewer photographed the completed diagram.

**Table 1. List of Core Constructs**

| | | |
|---|---|---|
| appropriateness of methodology | developer skill level | quality of software structure |
| appropriateness of programming paradigm | developer well-being | quality of system specifications |
| comprehension of software specifications | effectiveness of internal communication | quality of user involvement |
| consistency between specifications | financial risk | software complexity |
| cost/effort estimation accuracy | geographic distribution of work | team quality |
| degree of automation | management effectiveness | use of COTS[1] |
| degree of continuous improvement | measurability of software system | use of fault-tolerance mechanisms |
| degree of external uncertainty and change | quality assurance effectiveness | use of formal methods |
| degree of in-house reuse | quality of software requirements documentation | use of open source software |
| developer motivation | | |

[1]Commercial-Off-The-Shelf software

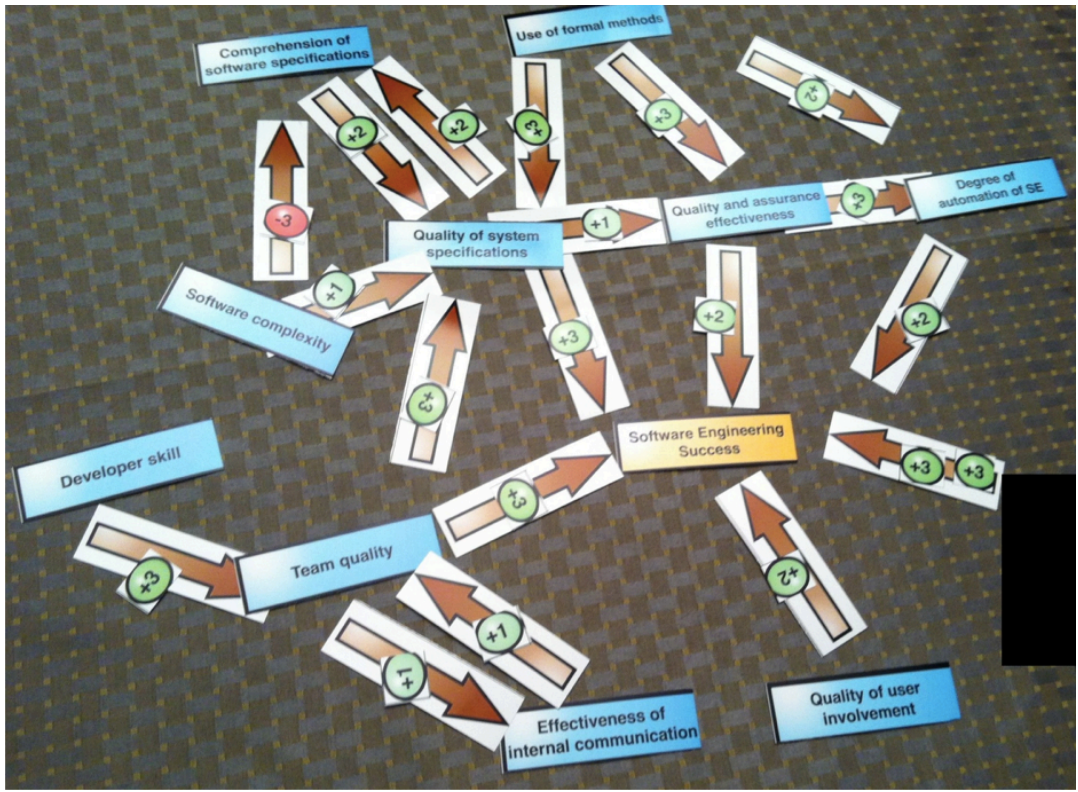

**Figure 1. Example of Causal Map**

*Notes: Blue rectangles are core constructs; arrows indicate causal direction; integers indicate magnitude of effect, integer sign indicates direct or inverse relationship; black rectangle obscures the participant's identity; yellow rectangle indicates the top level dependent variable.*

If a participant felt that an important construct was missing from the list, he or she had the option to write it on a blank card and add it. The following constructs were added this way: *being able to find artifact, architectural decisions, time to market, quality of process, customer involvement, suitability of organization culture, software architecture quality, problem, artifact evaluation.* No two participants added the same construct.

Participants could add causal arrows as they saw fit, as long as all constructs directly or indirectly affect SES. In practice, patterns included causal chains (e.g. *quality of system specifications* → *comprehension of system specifications* → SES), multiple effects (e.g. *management effectiveness* affects *team quality* and *developer well-being*), multiple antecedents (e.g. *management effectiveness* and *software complexity* affect *developer motivation*), reciprocal causation (e.g. *developer skill* affects *team quality* and vice versa) and loops (e.g. *management effectiveness* → *appropriateness of methodologies* → *quality of system structure* → *effectiveness of internal communication* → *management effectiveness*).

The use of *other* cards stems from a requirement of causal graph analysis. Strength of relationship is relative and constructs may be influenced by constructs not included in the diagram. Therefore, the relative strength of all *other* constructs is needed to understand the relative influence of included constructs. For *other* relationships, we recorded only relationship magnitude, as different other constructs may have opposite effects. In contrast, the absence of an *other* card for a construct indicates no other significant antecedents.

## 2.4. Hypothesis

As described in the introduction, Kuhn divided fields into three stages—1) no common ground, 2) competing schools with contradictory theories; 3) strong (but imperfect) majority agreement [21]. Taking an unsupervised cluster analysis approach (where the number of clusters is inferred from the data), we propose to interpret clusters as follows.

*Hypothesis $H_0$.* Clusters, if any, collectively include a minority of responses (no common ground).

*Hypothesis $H_1$.* Two or more clusters collectively contain the majority of responses (competing schools).

*Hypothesis $H_2$.* A single cluster contains the majority of responses (majority agreement).

Some interpretation may be required here. A single cluster encompassing 95% or more participants would clearly indicate stage 3, but a single cluster with 55%, or two clusters with 80% and 15%, would suggest that the community is transitioning from stage 2 to stage 3. Similarly, three clusters each with approximately 30% of participants would clearly indicate stage 2, while three clusters each having 15-20% with the remaining participants unclustered would suggest a middle ground between stage 1 and stage 2.

## 2.5. Methods of Analysis

We applied two clustering techniques—latent class analysis [9] and agglomerative hierarchical clustering [38]. Both methods classify the population completely into clusters. We therefore combined the results to distinguish robust clusters from incidental clusters.

Latent Class Analysis hypothesizes that a hidden variable, i.e. a latent class, is responsible for some of the variation of the observed population attributes. In our case, observed attributes are binary and denote whether a given construct is present in a graph or not, while the latent class represents which cluster the graph belongs to. Any variation beyond that caused by the latent class is attributed to chance. It also assumes that observed variables are independent of each other, except for the joint dependence on the latent class. These assumptions are captured in the following equation.

$$p(y, x_1 \ldots x_d) = q(y) \prod_{j=1}^{d} q_j(x_j, y)$$

where:
- $p(y, x_1..., x_d)$ denotes the probability that an individual graph belongs to cluster y
- $q(y)$ is the (unconditional) probability of class y, and
- $q_j(x,y)$ is the probability of attribute j assuming value x conditioned on y.

Using an expectation maximization maximum likelihood approach [38], the parameters $q(y)$ and $q(x,y)$ can be estimated to fit a given sample. With these estimates, each graph is assigned to the most probable class using the above equation, thus generating a cluster of graphs per class.

Maximum likelihood does not assist in the selection between model candidates with different numbers of latent classes. We therefore used the Akaike Information Criterion (AIC) [38] to identify the best candidate.

Clusters are identified based on some similarity. Causal graphs may be similar with respect to the included constructs and the existence, direction and strength of relationships between constructs. Latent class analysis uses only construct similarity; however, the Langfield-Smith Wirth causal map distance measure [22] uses all of these dimensions. We therefore based the agglomerative hierarchical clustering on the Langfield-Smith Wirth measure.

The agglomerative hierarchical clustering technique is based on an iterative, bottom-up approach. In the first iteration, the two closest causal maps according to the Langfield-Smith and Wirth measure are combined into a cluster. In the next pass, the second-closest pair is combined. The first cluster is included among the pairing candidates second pass. To determine the distance between clusters and individuals, or clusters and clusters, an aggregated score such as the maximum distance between any of the included individuals is calculated. The iterations continue until a single cluster remains. The end result is thus a hierarchy of clusters. More precisely, the Langfield-Smith Wirth causal map distance measure is defined as:

$$DR = \frac{\sum_{i=1}^{p} \sum_{j=1}^{p} |a_{ij}^* - b_{ij}^*|}{6pc^2 + 2pc(pu_1 + pu_2) + pu_1^2 + pu_2^2 - (6pc + pu_1 + pu_2)}$$

where:

$$a^*_{ij} = \begin{cases} 1, & \text{if } a_{ij} > 0 \text{ and } i \text{ or } j \notin P_c \\ -1, & \text{if } a_{ij} < 0 \text{ and } i \text{ or } j \notin P_c \\ a_{ij} & \text{otherwise} \end{cases}$$

- $b^*_{ij}$ follows a same pattern
- $a_{ij}$ is the value of causal map A's adjacency matrix element (i,j)
- $b_{ij}$ is the value of causal map B's adjacency matrix element (i,j)
- $P_c$ = the number of elements common to the two matrices
- $p$ = the number of elements in the adjacency matrix
- $pu_1$ = the number of unique elements in matrix A;
- $pu_2$ = the number of unique elements in matrix B.

## 3. FINDINGS

### 3.1. Data Description

Seven of the 28 constructs were selected by at least half of the participants (Table 2). However, only one relationship (between team quality and SES) was indicated by at least 50% of respondents (Table 3). While this does not necessarily imply low consensus, the lack of agreement around which constructs and relationships are most important is not what we would expect from a high-consensus field.

However, constructs may affect SES directly or indirectly, and to different degrees as participants indicated by numbering relationships. Frequency of inclusion in maps may therefore inaccurately indicate construct importance. A more accurate indication of importance is given by aggregate, transitive influence (Table 4), which indicates the strength of the effect of the construct on SES.

### 3.2. Clusters

The cluster analysis (Figure 2) reveals three robust clusters:

- Cluster A: graphs 23-32 and 48
- Cluster B: graphs 15-22, 45, 49-51, 53, 55, 57 and 58
- Cluster C: graphs 11, 13, 14 and 33-41

**Table 2. Most Popular Constructs (Excluding *other*)**

| Construct | % |
|---|---|
| developer skill | 80 |
| team quality | 72 |
| quality of user involvement | 72 |
| effectiveness of internal communication | 68 |
| degree of external uncertainty and change | 60 |
| developer motivation | 52 |
| software complexity | 50 |

**Table 4. Aggregated, Transitive Construct Influence[1]**

| Construct | % |
|---|---|
| team quality | 13 |
| quality and assurance effectiveness | 9 |
| quality of software structure | 8 |
| developer skill | 8 |
| degree of external uncertainty and change | 7 |
| software complexity | 6 |
| management effectiveness | 6 |
| quality of user involvement | 5 |
| degree of continuous improvement | 5 |
| effectiveness of internal communication | 4 |
| developer motivation | 4 |
| appropriateness of methodologies | 4 |
| quality of requirements documentation | 3 |
| comprehension of software specifications | 3 |
| cost/effort estimation accuracy | 3 |
| quality of system specifications | 2 |
| degree of automation of SE | 2 |
| measurability of software systems | 2 |
| consistency between specifications | 2 |
| financial risk | 2 |
| degree of inhouse reuse | 1 |
| use of formal methods | 1 |
| use of open source software | 1 |
| developer well-being | 1 |
| use of COTS | 1 |
| appropriateness of programming paradigm | 1 |
| geographic distribution of work | 0 |
| use of fault-tolerance mechanisms | 0 |

[1]The total influence exceeds 100% because many constructs due to their indirect influence are counted twice, so to speak.

**Table 3. Most Popular Relationships (Excluding *other*)**

| Cause | Effect | % |
|---|---|---|
| team quality | SES | 52 |
| developer skill | team quality | 38 |
| quality of software structure | SES | 38 |
| quality and assurance effectiveness | SES | 35 |
| degree of external uncertainty and change | SES | 30 |
| effectiveness of internal communication | team quality | 27 |
| developer skill | SES | 25 |
| software complexity | SES | 25 |
| quality of user involvement | SES | 23 |
| management effectiveness | team quality | 22 |
| degree of continuous improvement | SES | 20 |
| developer motivation | SES | 20 |
| developer motivation | team quality | 20 |

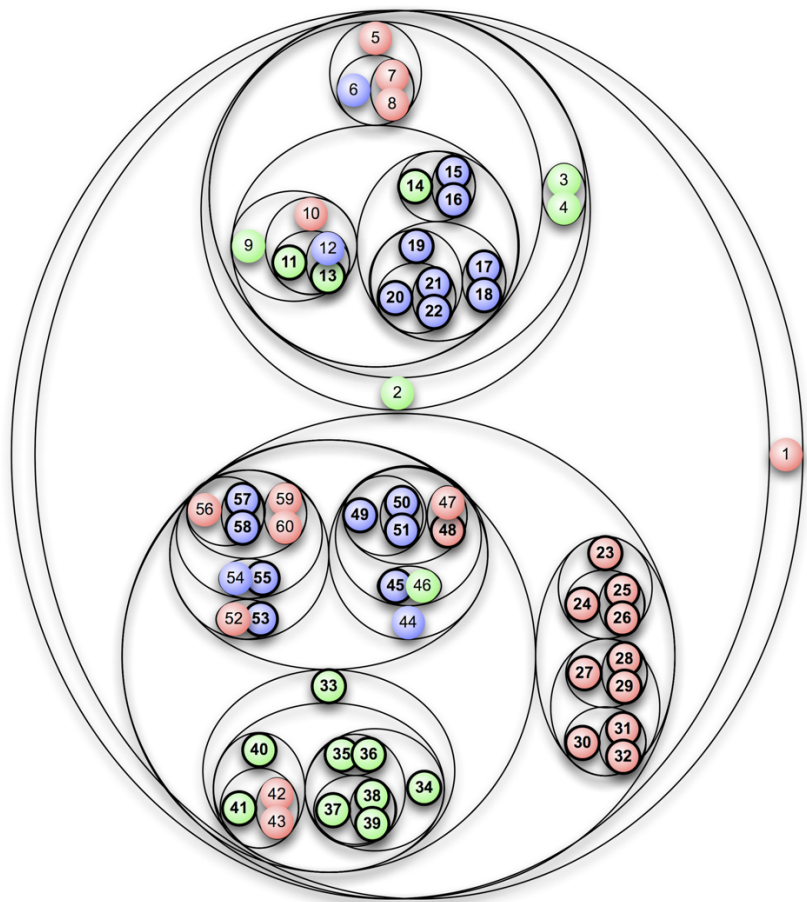

**Figure 2. Combined Cluster Analysis**

In Figure 2, each number represents a causal graph. Bold numbers represent causal graphs in a cluster while non-bold numbers are not in a cluster. Colors represent the three clusters resulting from the latent class analysis (pink for A, blue for B and green for C). Ellipses represent the clusters resulting from the agglomerative hierarchical clustering. Although the agglomerative hierarchical clustering generated more than three clusters, it does not generally find the clusters with the highest internal cohesion. Therefore, clusters can sometimes be extended without increasing the internal maximum Langfield-Smith and Wirth distance. In our case, extending the agglomerative hierarchical clusters to match the latent class clusters had no effect on cluster quality. This agreement between clustering algorithms supports the reliability of the three-cluster breakdown.

As most responses (39/60) are collectively contained in the three clusters, *Hypothesis $H_1$ is supported.* However, given that over a third of respondents are not clustered, we estimate that the software engineering community is somewhere between Kuhn's first stage (no consensus) and second stage (competing schools).

### 3.3. Consensus Graphs

The above analysis investigated the magnitude of consensus within SE. However, we are also interested in the nature of the consensus, i.e., exactly what did participants agree on? To investigate, we produced a *consensus graph* for each cluster and for the data overall. A consensus graph is an aggregation of individual construct maps highlighting the most agreed constructs and relationships.

Consensus graphs were created using the following procedure. For each construct in each graph, the squared weights of the antecedents, representing the correlation between the constructs as provided by the respondents, were normalized so that their combined effect summed to unity:

$$\sum_{\alpha \in A} \omega_\alpha^2 = 1$$

The fact that constructs and relationships are excluded from a graph also provides information about the beliefs of the respondent. The absence of a relationship between two included constructs indicates a very small positive or negative effect. The absence of a construct implies that its effect on SES is smaller than the least influential of the included constructs. This information is, however, in the form of ranges rather than as point estimates. As previously mentioned, the response alternatives for influences between constructs in the graph were limited to 0, 1, 2 or 3. The resulting approximation errors can also be represented as ranges (so that the response 1 corresponds to the range [0.5, 1.5]). A range:

$$[x_{i,min}^g, x_{i,max}^g]$$

was therefore used to represent the weight of each relationship, $x_i$, both those explicitly expressed by the respondent, g, and those whose values could be inferred. The relationships where no information could be inferred (e.g. between two constructs excluded from the graph) were assigned the full range corresponding to unnormalized values [-3.5, +3.5].

Individual causal graphs are then aggregated. The aggregate beliefs about the weight of a relationship can be represented as a function,

$$b(X_i = x_i) = \begin{cases} \sum_{g \in G} H_g, & \text{if } x_{i,min}^g < x_i < x_{i,max}^g \\ = 0, & \text{otherwise} \end{cases}$$

where:

- G is the set of graphs $\{g_1...g_m\}$, $x_i$ is an attribute, and

$$H_g = ln(\frac{1}{(x_{i,max}^g - x_{i,min}^g)})$$

is a measure of the entropy, moderating the impact of ignorance by attributing lesser importance to wide-ranged beliefs than to specific beliefs.

A measure of the aggregated belief per relationship was finally obtained by calculating the mean value of the above function. To generate a causal diagram of similar size in number of constructs and relationships as those collected, the top 99th, 98th and 97th percentiles of the most important relationships were extracted. These were then used as base for the diagrams, which are presented in Figures 3-6.

## 4. DISCUSSION

### 4.1. Interpretations and Implications

The above analysis suggests that SE is not a high-consensus field and therefore:

- Information created by the field for external stakeholders (e.g. standards, curriculum guidelines) do not reflect a consensus view because there is no consensus to reflect.
- SE research is substantially curtailed by academics unable to communicate across a paradigmatic divide or incrementally develop a cumulative body of knowledge.

Furthermore, this study investigates the degree of consensus but not the reasons for consensus. Indeed, higher levels of consensus are usually observed in "rapid-discovery fields" [10] including physics, chemistry, biology and climatology. SE appears to have a rapid pace of discovery. While dissensus may be dismissed as a consequence of SE's relative youth, three other differences between SE and many rapid discovery fields are evident

1. Unlike most rapid-discovery fields, SE includes considerable social and human aspects.
2. SE predominately deals with designing complex systems, which are innately unpredictable [16, 17].
3. Despite the great progress of empirical software engineering, SE does not manifest the strong commitment to empiricism seen in the natural sciences (cf. [26, 28]).

The above analysis further suggests that SE is characterized by at least three competing groups. All three groups included both management effectiveness and developer motivation. To highlight the unique aspects of each cluster, we cross-reference the constructs from the cluster consensus graphs (Figures 3-6) against the constructs that are popular overall (Table 3). Table 6 compares the three clusters showing unpopular constructs (that is, those peculiar to each cluster) in bold.

### 4.2. Validity and Limitations

The above results should be interpreted in light of the following limitations.

1. As the study used convenience sampling, we cannot statistically generalize the results to a larger population. However, a non-random sample is more likely to overestimate consensus than underestimate it.

The list of constructs may bias the results. The systematic manner in which constructs were chosen makes the study replicable but different lists may produce different results.

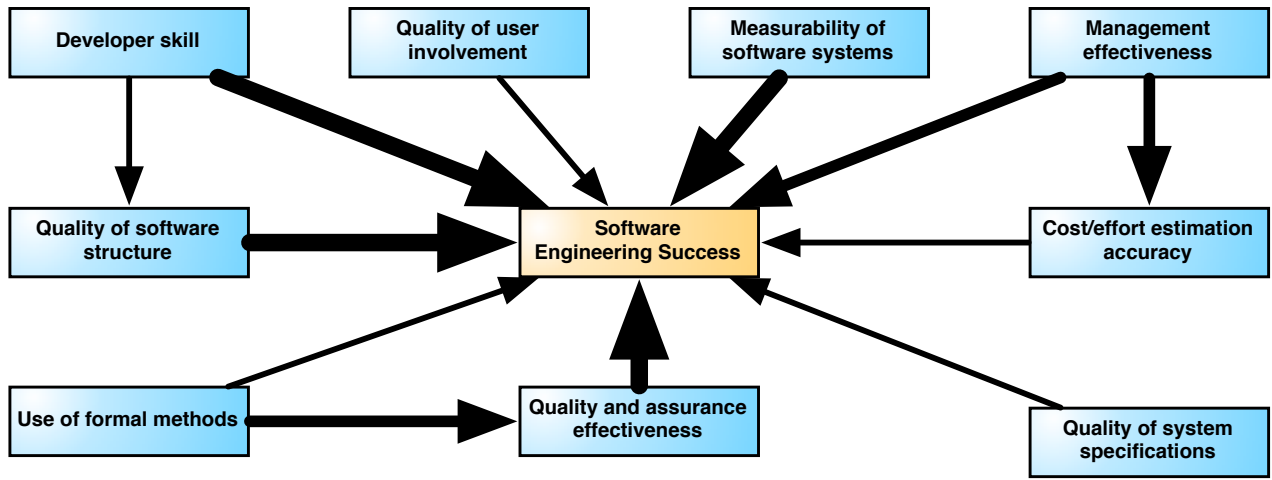

Figure 3. Consensus Graph – Cluster A

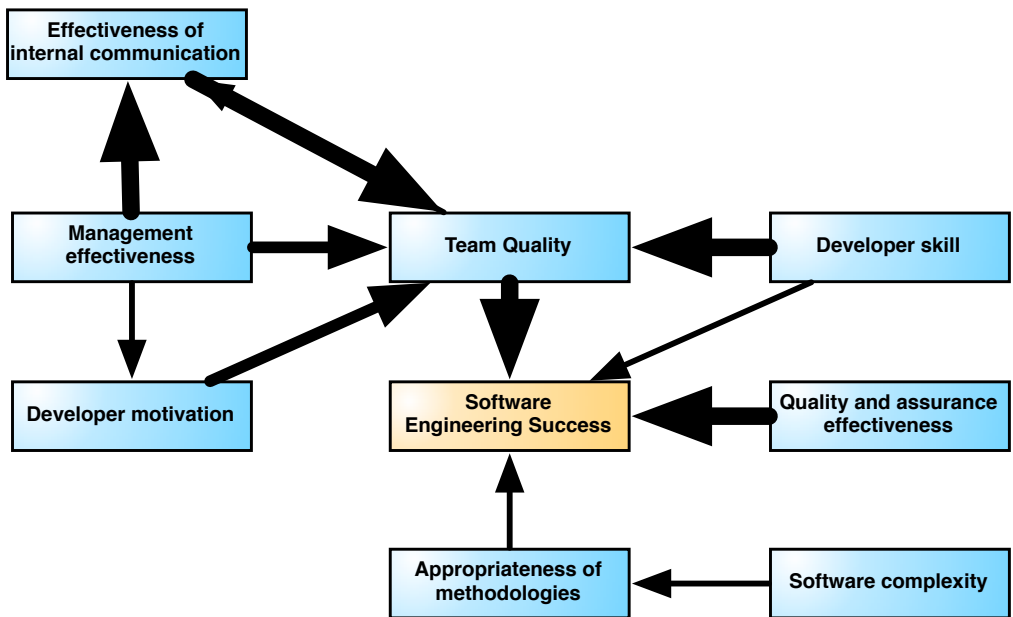

Figure 4. Consensus Graph – Cluster B

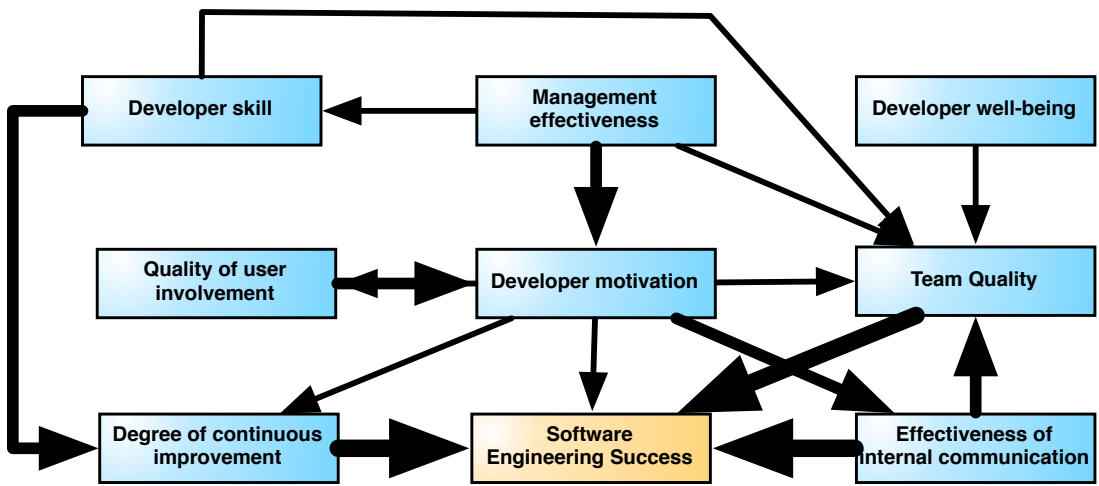

Figure 5. Consensus Graph – Cluster C

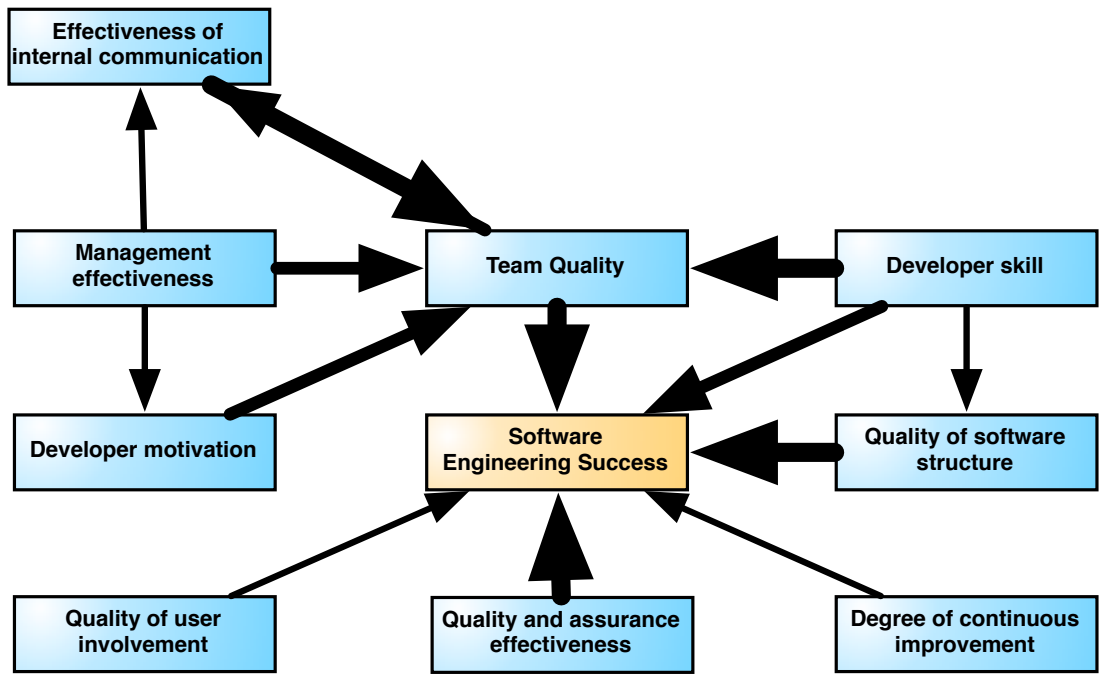

**Figure 6. Overall Consensus Graph**

**Table 6. Cluster Comparison[1]**

| Cluster A | Cluster B | Cluster C |
|---|---|---|
| **use of formal methods** | **appropriateness of methodologies** | **developer well-being** |
| **quality of system specifications** | software complexity | **quality of user involvement** |
| **cost/effort estimation accuracy** | effectiveness of internal communication | **degree of continuous improvement** |
| **measurability of software systems** | developer motivation | developer motivation |
| quality and assurance effectiveness | quality and assurance effectiveness | effectiveness of internal communication |
| quality of software structure | team quality | team quality |
| developer skill | developer skill | developer skill |
| management effectiveness | management effectiveness | management effectiveness |
| quality of user involvement | | |

[1]Unpopular constructs in bold

2. Participants could not represent non-linear relationships.
3. The effect of *other* in the causal maps may be exaggerated from our forcing it into the graphs or understated due to miserly information processing [37], i.e., participants not thinking through all of the possible *other* constructs.
4. Participants' causal maps may be biased by political motivations, e.g., mapping beliefs about how the constructs *should* relate rather than how they *actually* relate.
5. We investigated the implicit mental models of the participants, which do not necessarily correlate to the realities of software development. Again, our interest is in consensus objective causes of success.

More generally, this study uses success antecedents as a surrogate for overall consensus within the SE field. Put another way, we measure consensus in an SE subdomain (success antecedents) and theoretically generalize our results to SE overall. Clearly, lack of consensus around success antecedents does not infer lack of consensus on anything. However, selecting a particular domain was necessary to keep the study tractable. We chose success antecedents in particular as as previous work has highlighted agreement on software engineering success as the core top-level dependent variable for the SE field [33].

## 5. RELATED WORK
Four streams of related work are relevant:

1. the concept and importance of scientific consensus, which motivates this study;
2. causal and cognitive graphs, which inspire the methodology;
3. general theories of software engineering, which relate to the consensus graphs;
4. success factor frameworks, which are similar in some ways to the consensus graphs.

Scientific consensus has garnered increasing attention in epistemology, with several schools of thought. Positivists often hold that consensus generally characterizes science while postpositivists (e.g. [23]) claim that dissensus is more common. Natural sciences including physics and biology exhibit relatively high consensus while social sciences including sociology exhibit relatively low consensus [12, 21]. Kuhn [21] argued that scientific communities of practice generally begin with dissensus (pre-science). Later, several pre-paradigmatic schools of thought emerge. In some fields, one of the pre-paradigm schools triumphs over the competition, leading the field into consensual normal science. The new paradigm "must seem better than its competitors, but it need not, and in fact never does, explain all the facts with which it can be confronted" [21]. However, there is no guarantee that a scientific consensus accurately reflects reality or that new paradigms are more *accurate* than old ones [15]. New paradigms take hold when they are more useful for solving scientific puzzles, rather than when they are more accurate representations of reality [21].

Depending on one's particular philosophical position, the degree of consensus reflects either our level of knowledge about the real world, the maturity of research, or the social cohesion of the research community. However, scientific communities are not simply in consensus or dissensus. Within a field, individual propositions often begin at the research frontier, where dissensus dominates, and later move to the field's interior, where consensus dominates [23]. In fields characterized by rapid discovery, the advancing frontier drives consensus building by eclipsing bickering over yesterday's disagreements [10].

Meanwhile, causal maps (also called cognitive maps and causal cognitive maps) have long been used as a tool for exploring individual's beliefs and operationalizing tacit knowledge [34]. Methods for eliciting causal graphs are well-established and a growing body of research uses causal maps of groups of people [2, 20, 24]. Measures of similarity [22] and consensus [39] among multiple causal maps are used in many studies.

Furthermore, whether participants' beliefs about the constructs in this study reflect reality is an empirical question. However, from a theory-building perspective, we could reconceptualize the overall consensus graphs as an initial General Theory of Software Engineering (GTSE). Alternatively, we could reconceptualize the three cluster consensus graphs as competing GTSEs. Either thought experiment ties our results to existing GTSE research. Many papers have called for developing general or unifying theories in software engineering (e.g. [18, 19, 33]). Several initial attempts to generate a GTSE have emerged, alternatively based on transaction cost economics [13], design thinking [30, 31], SE as a social process [1] and SE as theory-manipulation [27]. Others (e.g. [29]) have argued for combining and adapting diverse theories from reference disciplines (e.g. psychology, sociology, management) to create a multi-level GTSE. However, even this multi-level view focuses more on SE processes and general causality (e.g. individual interactions cause team dynamics). The above consensus maps complement these processes theories by highlighting constructs peculiar to SE.

Moreover, numerous studies have investigated *critical success factors*, *risk factors* or *failure factors* in SE and related disciplines including project management and information systems (e.g. [3, 7, 8, 11, 14, 25, 35, 36]). A typical factor study proposes a taxonomy of constructs that are causally related to SE outcomes. Factor studies consequently suffer from two types of problems—1) substantial differences in taxonomy composition and structure between studies suggesting poor reliability; 2) most are based on ethnographic methods, which are intended to reveal the socially constructed perceptions of participants, not causal relationships between variables in an objective reality. In other words, many studies of success factors are actually studies of consensus. A key difference, however, is that we stress that our results reflect only consensus among participants (not reality) whereas many studies of success factors imply that their results are accurate depictions of reality.

# 6. CONCLUSION

This study has two main contributions. The above analysis indicates that the SE academic community of practice suffers from poor consensus. That is, some researchers are divided into several camps where each camp exhibits some internal consensus concerning the field's core phenomena, but significant disagreements separate the camps and many researchers are not in any camp. To the extent that any overall consensus exists, it seems to concern the importance of competence (developer skill, management effectiveness and team quality) for software engineering success. This consensus on competence is interesting and surprising since most SE research appears to focus on tools and technologies, rather than competence-building per se.

These conclusions suggest several avenues of future research. As suggested in the previous section, the consensus maps could be treated as theories and empirically tested, perhaps through longitudinal surveys or field studies. (Using further research based on expert opinion would invite common method bias). Meta-analysis or thematic synthesis of existing studies might also illuminate empirical support for many of the included relationships.

Another approach may involve extending constructs by integrating theories and models from reference disciplines. For instance, while this study found consensus around *team quality*, previous research in psychology and sports science has identified team cohesion as the primary determinent of team performance [4-6]. Problem solving, cognition, and decision theory are other examples of relevant disciplines. However, such empirical research is hampered by poor understanding of how to measure SES [32].

Finally, it bears repeating that scientific consensus is extremely important. Without a foundation of widely-agreed concepts and theories, SE may be uniformly dismissed by critics as unscientific—its standards and guidelines rejected as capricious. Consensus building is therefore critical to the field's legitimacy and survival. Consensus built on evidence rather than ideology would be all the better.

# 7. ACKNOWLEDGMENTS
Our thanks to the participants of this study for taking the time and effort to create their causal maps.

# 8. REFERENCES

[1] Adolph, S. and Kruchten, P. 2013. Generating a useful theory of software engineering. *Proceedings of the 2nd Workshop on a General Theory of Software Engineering*. International Conference on Software Engineering. 47–50.

[2] Ambrosini, V. and Bowman, C. 2001. Tacit knowledge: Some suggestions for operationalization. *Journal of Management Studies*. 38, 6 (2001), 811–829.

[3] Bavani, R. 2009. Critical success factors in distributed agile for outsourced product development. Proceedings of CONSEG-09, International Conference on Software Engineering.

[4] Beal, D.J., Cohen, R.R., Burke, M.J. and McLendon, C.L. 2003. Cohesion and Performance in Groups: A Meta-Analytic Clarification of Construct Relations. *Journal of Applied Psychology*. 88, 6 (2003), 989–1004.

[5] Bollen, K.A. and Hoyle, R.H. 1990. Perceived Cohesion: A Conceptual and Empirical Examination. *Social Forces*. 69, 2 (1990), 479–504.

[6] Carron, A.V., Bray, S.R. and Eys, M.A. 2002. Team cohesion and team success in sport. *Journal of Sports Sciences*. 20, 2 (2002/01/01 2002), 119–126.

[7] Charette, R.N.S.I. 2005. Why software fails. *Spectrum, IEEE*. 42, 9 (Sep. 2005).

[8] Chua, C. and Lim, W.K. 2009. The Role of IS Project Critical Success Factors: A Revelatory Case. *Proceedings of the International Conference on Information Systems*. AIS.

[9] Collins, L.M. and Lanza, S.T. 2010. *Latent class and latent transition analysis: With applications in the social, behavioral, and health sciences*. Wiley.

[10] Collins, R. 2001. Why the Social Sciences Won't Become High-Consensus, Rapid Discovery Science. *What's Wrong With Sociology?* S. Cole, ed. Transaction Publishers. 61–85.

[11] Cooke-Davies, T. 2002. The "real" success factors on projects. *International Journal of Project Management*. 20, 3 (2002), 185–190.

[12] de Solla Price, D.J. 1986. *Little science, big science... and beyond*. Columbia University Press New York.

[13] Erbas, C. and Erbas, B. 2013. An empirical approach to a general theory of software (engineering). *Proceedings of the 2nd Workshop on a General Theory of Software Engineering*. International Conference on Software Engineering. 23–26.

[14] Ewusi-Mensah, K. 2003. *Software Development Failures*. MIT Press.

[15] Feyerabend, P. 1975. *Against method: outline of an anarchistic theory of knowledge*. Humanities Press.

[16] Gell-Mann, M. 1999. Complex adaptive systems. *Complexity: Metaphors, models and reality*. Westview Press. 17–45.

[17] Holland, J.H. 1992. Complex Adaptive Systems. *Daedalus*. 121, 1 (Jan. 1992), 17–30.

[18] Johnson, P., Ekstedt, M. and Jacobson, I. 2012. Where's the Theory for Software Engineering? *IEEE Software*. 29, 5 (2012), 94–96.

[19] Johnson, P., Ralph, P., Goedicke, M., Ng, P.-W., Stol, K.-J., Smolander, K., Exman, I. and Perry, D.E. 2013. Report on the Second SEMAT Workshop on General Theory of Software Engineering (GTSE 2013). *SIGSOFT Software Engineering Notes*. 38, 5 (2013), 47–50.

[20] Kearney, A.R. and Kaplan, S. 1997. Toward a methodology for the measurement of knowledge structures of ordinary people: the conceptual content cognitive map (3CM). *Environment and Behavior*. 29, 5 (1997), 579–617.

[21] Kuhn, T.S. 1996. *The Structure of Scientific Revolutions*. University of Chicago Press.

[22] Langfield-Smith, K. and Wirth, A. 1992. Measuring differences between cognitive maps. *Journal of the Operational Research Society*. (1992), 1135–1150.

[23] Latour, B. 1987. *Science in action: How to follow scientists and engineers through society*. Harvard university press.

[24] Markíczy, L. and Goldberg, J. 1995. A method for eliciting and comparing causal maps. *Journal of Management*. 21, 2 (1995), 305–333.

[25] Misra, S.C., Kumar, V. and Kumar, U. 2009. Identifying some important success factors in adopting agile software development practices. *Journal of Systems and Software*. 82, 11 (Nov. 2009), 1869–1890.

[26] Parnas, D.L. 2003. The limits of empirical studies of software engineering. *Proceedings of the 2003 International Symposium on Empirical Software Engineering*. 2–5.

[27] Perry, D.E. 2013. A theoretical foundation for software engineering: A model calculus. *Proceedings of the 2nd Workshop on a General Theory of Software Engineering*. International Conference on Software Engineering. 39–46.

[28] Ralph, P. 2011. Introducing an Empirical Model of Design. *Proceedings of The 6th Mediterranean Conference on Information Systems*. AIS.

[29] Ralph, P. 2013. Possible core theories for software engineering. *Proceedings of the 2nd SEMAT Workshop on a General Theory of Software Engineering*. 35–38.

[30] Ralph, P. 2016. Software engineering process theory: A multi-method comparison of Sensemaking-Coevolution-Implementation Theory and Function-Behavior-Structure Theory. *Information and Software Technology*. 70, (2016), 232–250.

[31] Ralph, P. 2015. The Sensemaking-coevolution-implementation theory of software design. *Science of Computer Programming*. 101, (2015), 21–41.

[32] Ralph, P. and Kelly, P. 2014. The Dimensions of Software Engineering Success. *Proceedings of the International Conference on Software Engineering*. ACM. 24–35.

[33] Ralph, P., Johnson, P. and Jordan, H. 2013. Report on the First SEMAT Workshop on a General Theory of Software Engineering (GTSE 2012). *SIGSOFT Software Engineering Notes*. 38, 2 (2013), 26–28.

[34] Reber, A.S. 1989. Implicit learning and tacit knowledge. *Journal of Experimental Psychology: General*. 118, 3 (Sep. 1989), 219–235.

[35] Reel, J.S. 1999. Critical Success Factors In Software Projects. *IEEE Software*. 16, 3 (1999), 18–23.

[36] Shin, B. 2003. An Exploratory Investigation of System Success Factors in Data Warehousing. *Journal of the Association for Information Systems*. 4, (Jan. 2003), p141–168.

[37] Stanovich, K. 2009. *What Intelligence Tests Miss: The Psychology of Rational Thought*. Yale University Press.

[38] Tan, P.-N. 2007. *Introduction to data mining*. Pearson Education India.

[39] Tegarden, D.P. and Sheetz, S.D. 2003. Group cognitive mapping: a methodology and system for capturing and evaluating managerial and organizational cognition. *Omega - The International Journal of Management Science*. 31, 2 (2003), 113–125.